\documentclass[acmsmall]{acmart}

\usepackage{xcolor,subcaption,multicol,geometry,amsmath,amsthm,amsfonts,arydshln,pifont,algpseudocode,algorithm,xcolor,blkarray,enumitem,cleveref,bm,graphicx,multirow}
\usepackage{listings}
\usepackage{xcolor}

\setcopyright{cc}
\setcctype{by}
\acmJournal{PACMMOD}
\acmYear{2026} \acmVolume{4} \acmNumber{3 (SIGMOD)} \acmArticle{230}
\acmMonth{6} \acmDOI{10.1145/3802107}

\author{Zizhao Mo}
\affiliation{
	\institution{University of Macau}
	\country{Macau SAR, China}
}
\email{yc17461@connect.um.edu.mo}

\author{Junlin Chen}
\affiliation{
	\institution{University of Macau}
	\country{Macau SAR, China}
}
\email{yc57440@um.edu.mo}

\author{Huanle Xu}
\affiliation{
	\institution{University of Macau}
	\country{Macau SAR, China}
}
\email{huanlexu@um.edu.mo}
\authornote{Corresponding author}

\author{Chengzhong Xu}
\affiliation{
	\institution{University of Macau}
	\country{Macau SAR, China}
}
\email{czxu@um.edu.mo}


\received{October 2025}
\received[revised]{January 2026}
\received[accepted]{February 2026}

\graphicspath{{images/}{../images/}}
\crefname{section}{§}{§§}

\begin{document}
\title{Serving Hybrid LLM Loads with SLO Guarantees Using CPU-GPU Attention Piggybacking}
\begin{abstract}
Nowadays, service providers often deploy multiple types of LLM services within shared clusters. While the service co-location improves resource utilization, it introduces significant interference risks for latency-sensitive (LS) services—which have strict SLO requirements for inference latency—and severely constrains the service capacity of best-effort (BE) services due to limited available memory. To address interference, existing systems typically rely on reserving headroom to constrain BE resource usage. However, this approach's coarse granularity compromises the SLO compliance of the latency-sensitive service and unnecessarily restricts the generation potential of the best-effort service.

In this paper, we propose OmniServe, a novel LLM serving system that efficiently harnesses both CPU and GPU resources to mitigate interference and improve throughput. Central to OmniServe is the Attention Piggybacking mechanism, which effectively offloads the Attention computation of BE services to CPUs on the fly. This mechanism also facilitates asynchronous communication between CPU and GPU streams, preventing GPUs from being blocked while aggregating Attention results. Additionally, OmniServe incorporates a dynamic batching control policy to adapt to fluctuating request arrivals, facilitating Dense module computation using layer-wise batching. Experimental results show that OmniServe improves the SLO attainment rate for LS services by up to $1.48\times$ while enhancing BE serving throughput by up to $9.85\times$ compared to state-of-the-art systems.
\end{abstract}

\maketitle

\vspace{-0.5em}
\section{Introduction}
Nowadays, service providers aim to deliver a diverse range of services in datacenters using large language models (LLMs), to offer versatility and flexibility across various domains such as programming assistance and and question answering. Current foundational LLM models~\cite{gemini, gpt4o} are typically trained on vast amounts of real-world data, to capture knowledge from all aspects of human life.

While a variety of LLM services are deployed together in clusters to enhance resource utilization, their service requirements vary significantly. For example, online chatting is a latency-sensitive application~\cite{chatgpt} where users expect stringent service level objectives (SLOs) to ensure fast responses and high token generation rates. In contrast, many services focus on back-office tasks, such as benchmarking, form processing, and data wrangling~\cite{holistic, spreadsheetcoder, wrangle}. These workloads have more lenient SLOs and relatively low priority, requiring service provisioning on a best-effort basis. Furthermore, even within the same service type, non-uniform service provisioning for different users is a commonly adopted strategy. For example, service providers may offer free users best-effort access while ensuring rapid responses for paid users.

As LLM-based services become increasingly popular in data centers, it is critical to simultaneously optimize performance for different types of services~\cite{ibench, perfiso, firm}. This often requires maximizing generation throughput of best-effort (BE) service requests without negatively impacting SLO guarantees for latency-sensitive (LS) services. To achieve this goal, Llumnix~\cite{llumnix} proposes a priority-based scheduling strategy to dynamically schedule various requests across multiple LLM serving instances. For ensuring performance isolation between different types of services, Llumnix limits the KV cache usage for BE services by reserving headroom for LS services, recognizing the positive correlation between memory usage and the level of interference introduced to LS services.

Despite its operational simplicity in production environments, Llumnix's GPU memory-centric paradigm inadequately sustains balanced quality-of-service for hybrid workloads. First, its coarse-grained memory controls fail to effectively address compute resource contention—experiments reveal persistent 50\% LS latency variance even with fixed memory reservations for LS services (Fig.~\ref{fig:interference_decoding}(a)), attributable to unmanaged GPU SM/Memory bandwidth competition. Second, the framework’s rigid prioritization during LS traffic surges triggers BE starvation, reducing BE throughput significantly due to GPU memory monopolization.

Fortunately, our analysis reveals that the underutilized CPU resources in LLM serving clusters can be strategically leveraged to enhance BE computation while minimizing interference with LS services. Specifically, idle CPU cores can compute the offloaded BE Attention workloads to reduce contention on GPUs, while the abundant CPU memory accommodates significantly larger KV caches for BE requests. However, fully exploiting CPU for BE throughput faces a critical bottleneck: the 498.1$\times$ performance gap in Dense module computation between GPUs and CPUs. This disparity necessitates reserving Dense computations for GPUs, creating a dependency where GPUs must wait for Attention results computed on CPUs. To avoid GPU blocking, a robust synchronization mechanism is required to orchestrate heterogeneous compute streams across devices. This challenge is further compounded by the dynamic nature of LLM workloads. Fluctuating request arrival rates widen the CPU-GPU performance gap under larger batch sizes, making synchronization increasingly complex during traffic bursts.

To address these challenges associated with harnessing CPU resources, this paper proposes OmniServe, an efficient LLM system for serving hybrid loads with SLO-awareness. The first innovation of OmniServe is the introduction of Attention Piggybacking, a new CPU inference mechanism designed to effectively decouple CPU Attention computation from GPU inference. Specifically, Attention Piggybacking enables an asynchronous execution flow between CPUs and GPUs, allowing the GPU stream to perform inference without waiting for gathering immediate Attention results from the CPU stream. Dense computations for these overflowed BE requests can be opportunistically piggybacked later using layer-wise batching, introducing minimal interference to LS services with fluctuating request arrival rates. Unlike token-wise batching~\cite{orca}, the set of requests executed within a batch under layer-wise batching can vary across layers during a single token iteration.

The second innovation of OmniServe lies in the dynamic batching control through explicit latency quantification. OmniServe leverages the high predictability of service latency at the module level to dynamically determine the appropriate number of BE requests that can be piggybacked with LS inference using layer-wise batching during decoding. This approach not only ultimately ensures the SLO for LS services, but also enhances the BE service throughput.

The Attention Piggybacking mechanism and explicit batching control approach are highly flexible and can be seamlessly integrated with various parallelism techniques for distributed inference, as well as into state-of-the-art LLM serving systems, such as Prefill-Decode disaggregation~\cite{distserve, splitwise, exegpt} and chunk prefill~\cite{taming}. We have developed a prototype of OmniServe based on the vLLM framework~\cite{pagedattention}. Additionally, we conduct extensive experiments to evaluate OmniServe in a cluster consisting of one host with four A100 GPUs and four additional CPU-only hosts. All hosts features Intel Xeon(R) Gold 6342 CPUs. The results show that OmniServe can improve the SLO attainment rate for LS service by up to 1.48$\times$ while enhancing the BE serving throughput by up to $9.85\times$, compared to existing hybrid serving systems. To summarize, we have made the following contributions in this paper:

\begin{itemize}[leftmargin=*]
\item We conduct a comprehensive analysis to examine the interplay between BE and LS services, highlighting the need for a more efficient interference mitigation approach in hybrid serving.
\item We design a novel Attention piggybacking mechanism for efficient service co-location. With asynchronous CPU-GPU interaction, GPUs do not need to wait for CPU results, representing a fundamental departure from prior offloading approaches.
\item We introduce a dynamic scheduling policy that builds on the Attention piggybacking mechanism. Its core innovation is a fine-grained, layer-level batching strategy for processing LS and BE requests concurrently—a method fundamentally distinct from token-wise batching.
\end{itemize}

\section{Background and Motivation}
\label{sec:background}

\subsection{LLM Serving Basics}
\label{sec:basics}

\subsubsection{LLM inference workflow}

\begin{figure}[!t]
\centering

\includegraphics[width=0.56\linewidth]{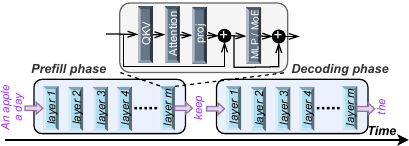}
\vspace{-.5em}
\caption{Illustration of the complete LLM inference workflow for token generation. All layers within the model must be executed sequentially and are composed of an identical set of computation modules.}
\label{fig:model_architecture}
\vspace{-1em}
\end{figure}

The LLM inference workflow, illustrated in Fig.~\ref{fig:model_architecture}, consists of two phases with distinct computational patterns: the \textit{Prefill} phase and the \textit{Decoding} phase. During the \textit{Prefill} phase (bottom-left), the entire input prompt is processed in a single, highly parallelizable forward pass to generate the initial output token (e.g., "keep"), making this a compute-intensive operation. The subsequent \textit{Decoding} phase (bottom-right) then generates tokens autoregressively, using the cumulative output sequence to produce one new token at a time (e.g., "the"). This phase is predominantly memory-bound, as each step requires loading the substantial KV cache for all previous tokens. Each token generation in both phases necessitates a complete forward pass through all sequential layers of the model. Notably, these layers share an identical structure, the components of which are detailed next.


\textbf{Dense computation}. In transformer-based models, Dense computation is responsible for capturing token-wise patterns. Specifically, each layer transforms the hidden states into $Q$, $K$, $V$ matrix in the \textit{QKV} module, translates Attention results in \textit{proj} module, and explores characteristics in a higher-dimensional space in \textit{MLP} module (replaced by the MoE module in some models). These operations are characterized by the high computational intensity, due to their matrix multiplication-based computation pattern. To this end, these modules are suitable to be executed on GPUs.

\textbf{Attention computation}. The Attention module plays a crucial role in LLM inference, enabling selective focus on specific parts of the context to highlight the most relevant information when generating responses \cite{attention}. Specifically, the Attention mechanism utilizes the representations of tokens—namely, the $Q$, $K$, and $V$ matrix—and computes Attention scores among them to capture their nuanced relationships. In decoding phase, the computation is formulated as:
\begin{equation}
\mathsf{Attention}(q, K, V)=\mathsf{softmax}\Big({q\cdot K^T}\big/{\sqrt{d_k}}\Big) \cdot V,
\end{equation}
where only the query $q$ of the last token is involved in computation with all previously stashed $K$ and $V$ matrices.

\textbf{Residual connection}. Residual connection provides an addition path for input tensors of each layer to bypass computations in the mainstream path~\cite{resnet}. This is presented in the upper part of Fig.~\ref{fig:model_architecture}, where the computational results of the \textit{proj} and \textit{MLP} modules have to \textit{add-and-normalize} with the activations stored before. Since this residual connection is able to enhance the numerical convergence, it has grown into the essential part of LLM models.

\subsubsection{LLM inference optimization}
To improve the inference performance, the following techniques are proposed:

\textbf{Continuous batching}. The continuous batching mechanism has recently been proposed to enhance resource utilization during the inference process~\cite{orca}. Specifically, it refines the batching granularity from the request level to the token level, allowing the number of requests in an inference batch that processes through all layers of the model to vary over time. This approach enables the saturation of computing resources for higher serving throughput, particularly in the decoding phase, which is less compute-intensive. To date, it has become a common practice in LLM inference \cite{pagedattention, lightllm, tgi}.

\textbf{Chunk Prefill}. The chunk prefill technique is proposed to process both prefill and decoding requests in a stall-free manner~\cite{taming}. To be specific, it divides a prefill request into chunks, which are then included in a batch alongside additional decoding requests. This approach can significantly reduce computational costs by avoiding the repeated invocation of the same kernel with identical model parameters, such as Dense modules, that would occur if the requests were processed in separate phases.

\subsubsection{LLM Inference Service Requirements}
Fast token generation is critical to LLM applications such as online chatting~\cite{chatgpt}, where users expect rapid responses from servers. Typically, the generation rate must meet a minimum threshold, i.e., the SLO constraint. Moreover, even for the same LLM service, varying service rates are expected across different phases. For instance, users are generally impatient to wait long for the generation of the first output token, while a relatively lower rate for subsequent tokens is acceptable due to limited human reading speed. To address these diverse token generation requirements, existing SLO-oriented systems like Splitwise and Distserve~\cite{splitwise, distserve} define key performance indicators for prefill and decoding phases: the Time to First Token (TTFT) and the Time Per Output Token (TPOT), respectively. In stark contrast, back-end inference tasks such as benchmarking~\cite{holistic}, form processing~\cite{spreadsheetcoder}, and data wrangling~\cite{wrangle} are of lower priority and do not require prompt responses. As such, no specific SLO requirement should be imposed for these LLM services. In this paper, we define LLM requests with SLO requirements as LS services and categorize others as BE services.

\begin{figure}[!t]
\begin{minipage}{0.2\linewidth}
\includegraphics[width=0.99\linewidth]{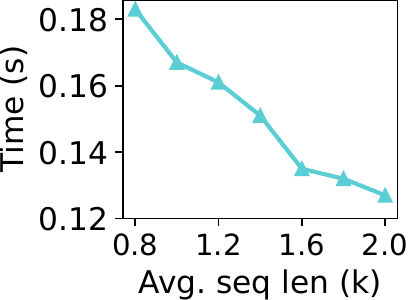}
\subcaption{Per-token latency}
\end{minipage}
\begin{minipage}{0.2\linewidth}
\includegraphics[width=0.99\linewidth]{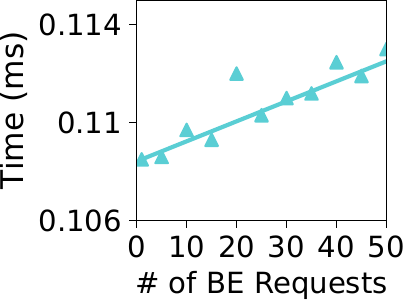}
\subcaption{MLP latency}
\end{minipage}
\begin{minipage}{0.2\linewidth}
\includegraphics[width=0.99\linewidth]{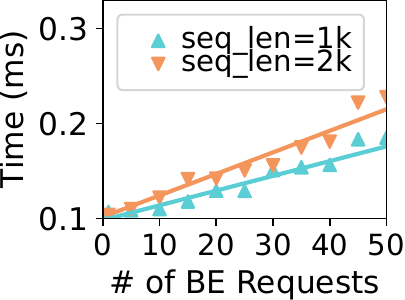}
\subcaption{Attention latency}
\end{minipage}
\vspace{-0.5em}
\caption{The latency of co-hosting LS and BE requests from Llama-70B models. (a): Significant per-token latency variation under Llumnix's fixed-size isolation, where the latency varies with the BE request length. (b) and (c): Latency of Attention and MLP modules given more BE requests.}
\vspace{-1em}
\label{fig:interference_decoding}
\end{figure}

\subsection{Existing Systems for  Hybrid LLM Serving Loads}
Recently, there exists a body of work focused on enhancing LLM serving performance from two key aspects - throughput~\cite{pagedattention, flexgen, orca} and latency~\cite{distserve, pagedattention, taming, shepherd, alpaserve, mark} - by effectively leveraging GPU or CPU resources. However, these approaches typically optimize for a single service type, overlooking the necessity to support multiple types of LLM services simultaneously. 
As a result, cluster administrators  are forced to deploy separate instances for different workloads, resulting in inefficient resource utilization. This approach duplicates model parameters in GPU memory and leaves compute resources in LS instances underutilized when BE requests could otherwise consume them.

Overcoming this resource waste necessitates multiplexing serving instances across both load types, complemented by delicate interference-aware resource management.
Llumnix~\cite{llumnix} is the first system that supports hybrid serving for LS and BE services. Specifically, it achieves runtime co-location by reserving dedicated cache space (headroom) for LS services, where the model parameters and KV cache spaces are shared by two services. In this sense, the idle resources of LS service at leisure time can be opportunistically utilized by the BE service, significantly enhancing the resource efficiency and improving the inference performance. However, this design is too coarse-grained and fails to guarantee SLO compliance, as it does not inherently address latency control mechanisms. We analyze this limitation and the interference in detail in~\cref{sec:interference}.

\begin{figure}[!t]
\centering
\includegraphics[width=0.5\linewidth]{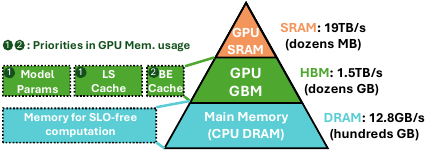}
\vspace{-0.5em}
\caption{Memory hierarchy along with the bandwidth and capacity information as well as the prioritization on memory resource usage, when LS and BE services are colocated on the same device.}
\label{fig:mem_contention}
\vspace{-1em}
\end{figure}

\subsection{Interference between LS and BE Services}
\label{sec:interference}
In this part, we investigate the bi-directional interference between LS and BE services, motivating the necessity of effective interference mitigation for hybrid workloads.

\subsubsection{Latency interference on LS service}
We analyze the impact of hosting additional BE requests on the decoding performance of LS services. Prior work has established that expanding the inference batch size increases latency due to resource contention, underscoring the necessity of latency control mechanisms for co-locating BE and LS requests~\cite{distserve, orca, pagedattention}. However, our experiments reveal that Llumnix’s headroom-oriented technique alone fails to sufficiently mitigate SLO violations. For instance, when deploying a 4-way~\footnote{4-way TP means the LLM inference instance is parallelized on four GPUs in tensor-parallel manner.} tensor-parallel (TP) Llama-70B model on four A100 80GB GPUs with 80\% of the KV cache reserved for LS requests, we observe a 1.47$\times$ increase in per-token latency for LS decoding under concurrent BE workloads compared to exclusive serving (Fig.~\ref{fig:interference_decoding}(a)). This degradation varies with different BE context lengths. Notably, interference worsens when the same cache space is allocated to prefill-phase BE requests, as their computationally intensive operations exacerbate contention.

\subsubsection{Serving capacity reduction on BE service} 
In hybrid load serving, LS services take precedence as the highest priority, inevitably impacting the performance of co-located BE services. Specifically, the number of BE requests that can be processed on GPUs is constrained by the limited memory available to BE services. As illustrated in Fig.~\ref{fig:mem_contention}, modern GPUs typically provide only tens of gigabytes of memory. After allocating space for LS service components—such as model parameters and dedicated caches—the remaining memory left for BE services becomes insufficient. This results in computational under-utilization: despite maintaining the SLO of LS services, the GPU’s computational resources such as SM cores cannot be fully leveraged by BE workloads due to memory constraints.

\subsection{Hybrid Serving: Opportunities and Challenges}

\subsubsection{Opportunities}
While multiplexing LS and BE services can possibly introduce significant interference, we identify opportunities that can be leveraged to ensure SLO compliance for LS services while enhancing BE throughput.

\textbf{$\bm{O_1}$: Harnessing ample CPU resources for BE computation.} CPU resources in LLM serving clusters are typically underutilized, as only the control flow for GPU workers is handled by the CPU~\cite{zhang2023cocktailer}, leaving the majority of cores idle during inference (e.g., the Intel Gold 6342 CPU has 24 physical cores, yet only four are actively utilized when running a 4-way inference instance)~\cite{pagedattention, lightllm, sglang}. Consequently, offloading the Attention computation of BE requests to the CPU can mitigate interference for LS requests. This strategy is grounded in empirical analysis: BE Attention modules—unlike their MLP counterparts—disproportionately degrade LS service performance due to intense contention for memory bandwidth and SRAM resources during decoding, as validated in Fig.~\ref{fig:interference_decoding}(b-c). Concurrently, leveraging CPUs’ abundant memory to host larger KV caches for BE requests (Fig.~\ref{fig:mem_contention}) reduces GPU memory pressure, enhancing BE throughput by minimizing contention while preserving LS performance.

\textbf{$\bm{O_2}$: Leveraging high predictability of service latency.} Due to the layer-wise organization of LLM and the uniform GEMM-based implementation of Dense modules, the whole computation process during inference is highly deterministic for specific inputs of given sizes, allowing for precise latency estimation of inference requests. This accurate estimation enables a reliable quantification of BE requests that can be multiplexed with LS services to ensure SLO guarantees. Additionally, the predictability of latency facilitates fine-grained load control, allowing BE requests to maximize the utilization of available GPU and CPU resources. Notably, the nearly consistent execution time observed with increased loads on MLP computation (illustrated in Fig.~\ref{fig:interference_decoding}(b)) enhances BE computation, thereby improving overall serving throughput.

\subsubsection{Challenges}
Despite the potential opportunities to benefit BE services while providing SLO guarantees for LS services, there are still several fundamental challenges when we want to fully utilize the CPU resources. To illustrate this, we conducted a series of experiments in our cluster, measuring inference performance for requests on an Intel Xeon Gold 6342 CPU and an NVIDIA A100 GPU, respectively. Our analysis focused on the CPU-GPU performance gap across different model modules (Attention versus MLP), operational phases (prefill versus decode), and batch sizes (1 versus 10). Specifically, the following key challenges were observed:

\textbf{$\bm{C_1}$: Low resource efficiency in partitioning computation between CPU and GPU.} As evidenced by Table~\ref{tab:gap}, a substantial performance gap exists between MLP execution on CPUs and GPUs—even for single-request decoding. This renders offloading MLP computations—including partial-layer implementations (e.g.,~\cite{llamacpp, pagedattention})—to CPUs ineffective for throughput improvement. However, existing inference systems’ reliance on token-wise continuous batching~\cite{orca} forces all batched requests to stall their post-Attention Dense modules (e.g., \texttt{proj} in Fig.~\ref{fig:model_architecture}) until all Attention results, including those computed on CPUs, are synchronized. This tight CPU-GPU coupling risks blocking GPU execution due to the CPU’s limited bandwidth and computational capacity, thereby prolonging LS service latency. Meanwhile, CPU resources remain underutilized while Dense modules execute on GPUs, further exacerbating cluster inefficiency.

\textbf{$\bm{C_2}$: Offloading computation in the presence of fluctuating inference loads.} The dynamic nature of LLM inference, characterized by unpredictable request arrivals, results in highly fluctuating serving loads. This variability further exacerbates the challenges posed by the performance gap between CPUs and GPUs. Specifically, the computation gap in the Attention module between CPU and GPU increases from 2.34$\times$ to 7.58$\times$ as the batch size of requests rises from 1 to 10, as shown in Table \ref{tab:gap}. This disparity significantly hampers the ability to ensure SLO guarantees for LS services, as the involvement of CPUs can easily degrade the token generation rate during peak BE loads due to poor synchronization with GPUs.  Consequently, it is crucial to implement a load-adaptive offloading scheme, coupled with a meticulous batching strategy for all requests with dynamic arrivals.

\begin{table}[t]
\small
\centering
\caption{The computation power gap across modules in the Llama-2-70B model between an Intel Xeon Gold 6342 CPU and an A100 GPU. The length of the decoding context and prefill request is both 1000 in the experiment}
\label{tab:gap}
\begin{tabular}{|c|cc|cc|}
\hline
\multirow{2}{*}{} & \multicolumn{2}{c|}{Prefill}          & \multicolumn{2}{c|}{Decode}         \\ \cline{2-5} 
                  & \multicolumn{1}{c|}{Attention} & MLP & \multicolumn{1}{c|}{Attention} & MLP \\ \hline\hline
1 request & \multicolumn{1}{c|}{$184.6\times$}         & $288.2\times$   & \multicolumn{1}{c|}{$2.34\times$}         & $65.2\times$   \\ \hline
10 requests & \multicolumn{1}{c|}{$393.75\times$}         &  $212.1\times$   & \multicolumn{1}{c|}{$7.58\times$}         & $498.1\times$   \\ \hline
\end{tabular}
\vspace{-1em}
\end{table}


\section{OmniServe System}
\subsection{Overview of OmniServe}
\subsubsection{Key ideas}
OmniServe is a novel LLM serving system designed to optimize performance for both LS and BE requests, with the capability to fully leverage CPU and GPU resources. To be specific, it is built on the following key design ideas:

\textbf{$\bm{I_1}$: Integrating the Attention Piggybacking mechanism.} OmniServe introduces the Attention Piggybacking mechanism, a novel design to enhance the performance of hybrid services across heterogeneous hardware. By decoupling CPU and GPU computation streams, this mechanism enables asynchronous execution: the GPU stream progresses with inference tasks without stalling for immediate Attention results from the CPU. Overflowed BE requests have their Dense computations (e.g., MLP) opportunistically piggybacked onto the GPU via a layer-wise batching technique once the dependent Attention outputs are available. As a consequence, the CPU stream focuses exclusively on processing Attention computations for these overflowed BE requests, while the GPU prioritizes executing the full generation pipeline for as many requests as possible. This symbiotic design ensures both CPUs and GPUs operate at peak efficiency for their respective workloads, directly addressing Challenge $\bm{C_1}$.

\textbf{$\bm{I_2}$: Dynamic batching control through explicit latency quantification.} To ensure SLO for LS services amid varying computational demands during the prefill and decoding phases, OmniServe proposes a module-wise latency modeling approach in witnessing the inconsistent latency increase across modules (Fig.~\ref{fig:interference_decoding}(b-c)). By quantifying these effects, OmniServe enables the  delicate piggybacking of Attention results to perform subsequent Dense computations using layer-wise batching. 
As a result, the throughput of BE generation is improved while still guaranteeing SLO for LS services, particularly in the face of fluctuating LS serving loads, thereby addressing Challenge $\bm{C_2}$.

\begin{figure}
\centering
\includegraphics[width=0.5\linewidth]{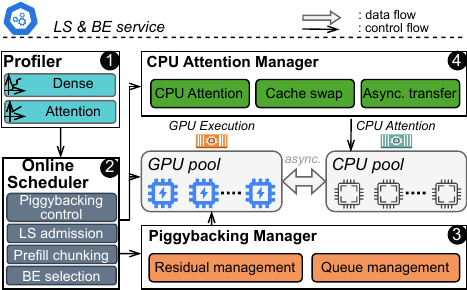}
\vspace{-0.5em}
\caption{The system overview of OmniServe.}
\label{fig:architecture}
\vspace{-1em}
\end{figure}

\subsubsection{System architecture}
The system architecture of OmniServe is illustrated in Fig.~\ref{fig:architecture} and comprises the following four main components: Profiler~\ding{182}, Online Scheduler~\ding{183}, Piggybacking Manager~\ding{184}, and CPU Attention Manager~\ding{185}. 

When OmniServe is deployed within an LLM serving instance, the Profiler initializes the modeling process to create precise performance profiles for various modules, supporting dynamic inference across CPU and GPU resources. The Online Scheduler determines the execution of requests on GPU and sustains the asynchronous CPU-GPU piggybacking based on SLO requirements. Subsequently, leveraging analytical models for Dense and Attention computations, the Online Scheduler dynamically chooses an appropriate number of LS and BE requests for execution within a batch. It also decides the number of tokens to be piggybacked from CPUs using layer-wise batching. The online scheduling functionality is further facilitated by the Piggybacking Manager, ensuring seamless and efficient Attention Piggybacking within the GPU backend through efficient activation management and queue design. To fully harness resources for BE services, excessive requests access KV cache on GPUs and swap them to memory, process intermediate results, and execute Attention with support from the CPU Attention Manager.

\subsection{Attention Piggybacking}
\label{sec:piggybacking}

The core rationale behind integrating Attention Piggybacking stems from the limitations associated with the continuous batching approach~\cite{orca}. Specifically, continuous batching inherently aligns the inference processes of BE and LS requests within a single batch at the token granularity, running them on the same hardware concurrently. This batching strategy hinders the effective utilization of either CPU resources or GPU resources due to a substantial performance gap in Dense module computations and the stringent SLO demands of the LS service. In this part, we illustrate the Attention Piggybacking mechanism, encompassing Attention offloading and post-Attention Piggybacking, to demonstrate how this approach benefits both BE and LS services simultaneously.

\begin{figure}[!t]
\centering
\includegraphics[width=0.45\linewidth]{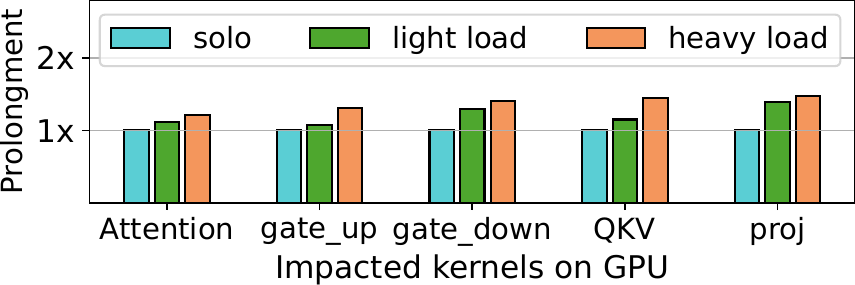}
\vspace{-.5em}
\caption{The interference of colocating kernel execution within the same CUDA context is as high as $1.5\times$ within a Llama-2-70B instance on A100 GPUs, motivating the importance of post-Attention piggybacking.}
\label{fig:colocation}
\vspace{-1em}
\end{figure}

\subsubsection{Attention offloading}
The first characteristic of Attention Piggybacking lies in the Attention offloading design, which involves selectively offloading the Attention computation to CPUs while keeping computations of other modules on GPUs. This design is depicted in the left portion of Fig.~\ref{fig:interaction}(a). In this example, the input tensor of the $l-$th layer consists of the data from three decoding LS requests and three decoding BE requests. Before this iteration, all the KV caches of these three BE requests have been offloaded to the CPU due to GPU memory shortage. As such, after the \texttt{QKV} computation is completed, the intermediate tensor of these three BE requests is forwarded to the CPU to proceed with their Attention computation there, and the LS requests continue their subsequent Attention computation on the GPU. Unlike synchronous execution flows, where GPU streams must wait for results from lower-end processors, OmniServe decouples the computation of Dense modules among LS and BE services on GPU devices without applying the token-wise continuous batching, i.e., no synchronous gathering after the Attention execution. This prevents the GPU's execution from being hindered by dependencies on immediate result gathering, particularly under the non-negligible data transfer latency given the limited PCIe bandwidth.

\subsubsection{Post-Attention Piggybacking} 
Attention Piggybacking, which defers immediate aggregation between LS and BE requests, inevitably raises a critical concern, i.e., when to transfer the CPU-computed Attention result back to GPU and how to proceed with the subsequent computations. 

To mitigate interference in the computation of LS requests, one potential approach is to block the kernel for offloaded BE requests until the GPUs become idle. However, this method may prevent BE requests from effectively utilizing GPU resources, which contradicts the goal of improving utilization for BE workloads. Alternatively, another strategy is to run both LS and BE inference workflows concurrently on the GPUs. While this approach allows for instantaneous BE execution, it introduces significant interference to LS services. In Fig.~\ref{fig:colocation}, we analyze the interference on various computation modules within Llama-2-70B when concurrently invoking a \texttt{proj} module on A100 GPU, which is the function following the Attention module. MLP module is divided into \texttt{gate\_up} and \texttt{gate\_down}. Each impacted module performs computation for 50 requests, each with a context length of 500. `Light' and `heavy' loads refer to scenarios where the \texttt{proj} kernel carries 5 and 200 requests. We isolate the BE and LS workflows on the GPU using two different CUDA streams to enable parallel execution. Results show that triggering the \texttt{proj} computation can uniformly disrupt the GPU workflow, depending on the other kernel invoked for LS service inference. Even with light loads on the \texttt{proj} kernel, handling only 5 requests, we observe slowdowns ranging from $1.12\times$ to $1.3\times$ for the concurrently running kernel. This interference can increase to $1.5\times$ when processing 200 requests within the \texttt{proj} kernel. The primary reason for this high interference is the substantial loading of model parameters for Dense computation, intensifying contention on GPU resources like SRAM and memory bandwidth. While the Attention kernel of LS services experiences minimal interference due to its low computational demand and parameter-free nature~\cite{attention}, its subsequent computation still encounters interference.

\begin{figure*}
\centering
\begin{minipage}{0.62\linewidth}
\includegraphics[width=0.99\linewidth]{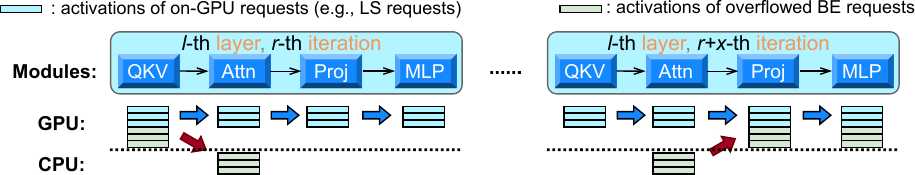}
\subcaption{Attention offloading and piggybacking design}
\end{minipage}
\begin{minipage}{0.33\linewidth}
\includegraphics[width=0.99\linewidth]{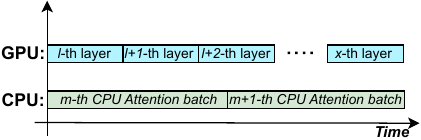}
\subcaption{Temporal view of CPU and GPU execution}
\end{minipage}
\vspace{-1em}
\caption{The Attention piggybacking mechanism within OmniServe. (a) Module-Level Dataflow: The design includes Attention offloading (left) and post-Attention piggybacking (right). Activation values of overflowed BE requests are offloaded to the CPU before executing the Attention module, while the resulting Attention outputs are aggregated back into the GPU dataflow after the same layer’s Attention module. (b) Temporal Workflow View: The CPU and GPU are utilized for diverse computations at arbitrary times, enabling efficient resource utilization and minimizing idle periods.}
\label{fig:interaction}
\vspace{-1em}
\end{figure*}

OmniServe addresses this interference by leveraging the recurrence of kernel invocation during the token generation process. As illustrated in the right part of Fig.~\ref{fig:interaction}(a), the \texttt{proj} module, located after the Attention module, is repeatedly invoked throughout the auto-regressive generation process. This recurrence provides an opportunity to piggyback the computation of offloaded BE requests within the subsequent Dense modules, minimizing interference with LS requests. Coupled with the asynchronous Attention offloading mechanism, this design allows the CPU and GPU to handle diverse computations in parallel, as shown in Fig.~\ref{fig:interaction}(b), thereby achieving full utilization of computational resources.

\subsubsection{Asynchronous CPU-GPU interaction}
\begin{figure}[t!]
\centering
\includegraphics[width=0.5\linewidth]{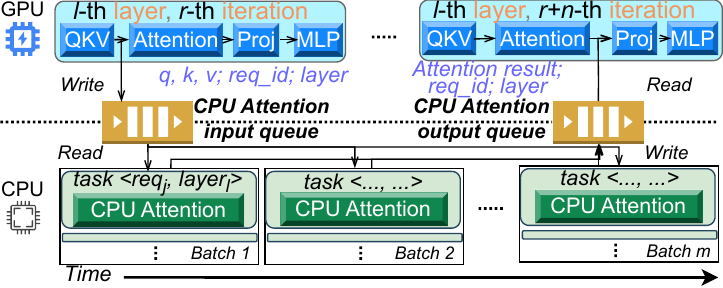}
\vspace{-.5em}
\caption{The CPU–GPU interaction in OmniServe, implemented using CPU Attention Input Queue and the CPU Attention Output Queue. This queue-based design enables asynchronous, interference-free communication and prevents LS service on the GPU from being affected by PCIe latency.
}
\vspace{-1em}
\label{fig:queue}
\end{figure}

The Attention piggybacking mechanism relies on the an efficient CPU-GPU interaction design, as illustrated in Fig.~\ref{fig:queue}.  Specifically, upon completion of the \texttt{QKV} module computation for both LS and BE requests at layer $l$ during the current iteration, the corresponding $q$, $k$, and $v$ vectors of BE requests are transferred from GPUs to CPUs for subsequent Attention computation. This communication is lightweight since only the last token of each request is involved in this process. Moreover, the KV cache transfer between the CPU and GPU is a one-shot operation for each request. As such, the communication over PCIe will not become a performance bottleneck.

As the GPU stream advances, when it is time to re-execute the \texttt{proj} module at layer $l$ after a series of token iterations, and the CPU has returned the Attention result of the $l$-th layer, the \texttt{proj} computation is performed using layer-wise batching on a modified input tensor that concatenates the returned Attention result from the CPU. Notably, adding additional BE decoding requests during the computation of Dense modules has minimal impact on LS requests, as depicted in Fig.~\ref{fig:interference_decoding}(b). This is because computations from both service types can be processed in one GeMM kernel. As CPU Attention results are transferred to the GPUs during GPU kernel execution, effectively hiding the transfer overhead.

To facilitate efficient data interaction between the CPU stream and GPU stream, the Piggybacking Manager incorporates two distinct queues for storing CPU Attention input and output. Within these queues, computed results are dispatched to the tail of the respective queue by the producer, while consumer retrieves data from the head of the queue, establishing a producer-consumer pattern. Specifically, the CPU-side write and read operations are managed by the CPU Attention Manager, while the GPU-side write and read operations are handled by the Piggybacking Manager.
According to queuing theory, in a stable state, the arrival rate of the queue equals the departure rate.  This indicates that the CPU stream and GPU stream will maintain a steady pace during computation, under this asynchronous execution paradigm. As a result, both the CPU and GPU resources can be fully utilized. Consequently, with more idle CPU resources available in data centers, they can store additional KV caches and execute Attention computation more rapidly through token-level parallelization, thereby substantially boosting the BE serving throughput. This queue-based design benefits more complicated serving scenarios, such as handling CPU failure and offloading the KV cache to disk. This is because it decouples the serving workflow from the cache source, enabling smooth execution in the event of loading from disk or abrupt fault occurrence.

\subsubsection{KV cache management}
\label{sec:cache_management}
Co-hosting BE and LS requests significantly increases memory usage in clusters. As a result, frequent KV cache swapping between CPUs and GPUs for BE requests may occur in response to fluctuating LS serving loads, leading to excessive KV cache migration that can degrade overall service performance. To address this issue, the CPU Attention Manager introduces an asynchronous KV cache swapping mechanism to manage memory swapping for BE requests. Specifically, it supports a non-blocking swapping-out operation by overlapping computation and data transfer on the GPU, preventing interruptions to LS inference. Additionally, the CPU Attention Manager implements a delayed swapping-in scheme for BE requests. When LS loads are light, caches for BE requests are not swapped in GPU memory immediately; instead, this process is triggered only after the $k$ and $v$ vectors of the last token are generated for all layers. This delay helps mitigate excessive cache swapping over the PCIe channel, which can occur when LS loads fluctuate wildly.

\subsection{Online Scheduling for Hybrid Loads}
\label{sec:scheduling}
\subsubsection{Inference latency models}
\label{sec:modeling}

Accurate performance modeling is essential for maintaining consistent performance isolation and maximizing the benefit from adopting the attention piggybacking design on the fly. However, existing modeling techniques rely on an assumption that the input and output tensors for modules in all model layers would have the identical size, rendering them unsuitable for our new Attention piggybacking mechanism. To achieve this, the Profiler incorporates novel analytical models with high-generality to estimate the execution time among Dense and Attention modules, represented by $f_D(\cdot)$ and $f_A(\cdot)$, collectively constituting the inference latency.

\begin{figure}[t]
\begin{minipage}{0.2\linewidth}
\includegraphics[width=0.99\linewidth]{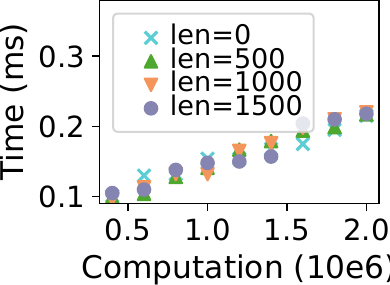}
\subcaption{Prefill Attention}
\end{minipage}
\begin{minipage}{0.2\linewidth}
\includegraphics[width=0.99\linewidth]{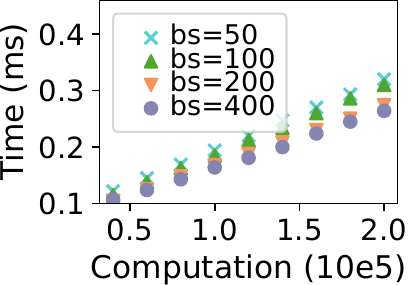}
\subcaption{Decode Attention}
\end{minipage}
\begin{minipage}{0.2\linewidth}
\includegraphics[width=0.99\linewidth]{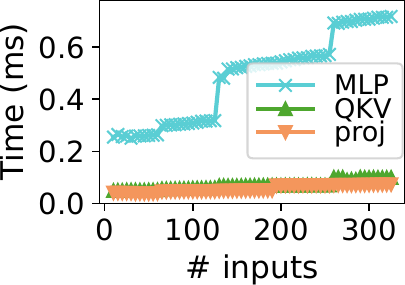}
\subcaption{Dense modules}
\end{minipage}
\vspace{-0.5em}
\caption{Attention and Dense computation characterization in 4-way TP Llama-2-70B model. (a) Prefill Attention time is only correlated to computation load. (b) Decoding Attention time is related to computation load and number of requests. (c) Dense computation time increases in a ladder pattern. }
\vspace{-1em}
\label{fig:modeling}
\end{figure}

\textbf{Modeling GPU Attention computation latency.}
Since requests from different computational phases can be served within a single batch, the latency model must account for both prefill and decoding attention simultaneously—a capability widely supported by advanced libraries~\cite{pagedattention, taming, xformers, flashinfer}. However, effective modeling remains challenging due to the presence of multiple variables that can influence the latency, such as the prompt and context lengths of each request. Fortunately, the pairwise token relationship evaluation (where causal masking governs token interactions in auto-regressive generation) enables a unified framework for latency modeling across diverse operational scenarios. By systematically characterizing latency variations as a function of Attention computational intensity, we demonstrate that prefill-phase Attention latency scales linearly with computation loads, remaining independent of input sequence lengths during chunked prefill operations, as shown in Fig.~\ref{fig:modeling}(a). Consequently, the Profiler models the prefill Attention latency $f_{PA}(\cdot)$ with learnable parameters $\{a_{PA},\ b_{PA}\}$ as:
\begin{equation}
f_{PA}\big(c_{PA}(t)\big)=a_{PA}\cdot c_{PA}(t) + b_{PA}.
\label{eq:attention}
\end{equation}
Here, for a prefill request $j$ with $l_j$ tokens already processed and $q_j$ tokens to be prefilled at time $t$, the Profiler models the computation load as $c_{PA}(t)=\sum_{i=l_j+1}^{l_j+q_j} i$.

During the decoding phase, only the latest token of each request requires attention computation. This allows the computational load to be simplified as: $c_{DA}(t)=\sum_{j\in g(t)} l_j+1$, where $g(t)$ denotes the number of decoding requests. Due to reduced computational intensity, decoding attention becomes memory bandwidth-bound, enabling higher compute utilization through parallel processing of multiple requests. As shown in Fig.~\ref{fig:modeling}(b), latency improves with increasing $g(t)$, even when the total KV cache size remains constant. To model this behavior, the Profiler defines the decoding attention latency $f_{DA}(\cdot)$  using learnable parameters $\{a_{DA},\ h_{DA},\ b_{DA}\}$:
\begin{equation}
f_{DA}\big(c_{DA}(t), g(t)\big)=a_{DA}\cdot c_{DA}(t) + h_{DA}\cdot g(t) + b_{DA},
\label{eq:attention}
\end{equation}
where the $h_{DA}\cdot g(t)$ term is used to capture the impact of compute utilization on latency. Combining $f_{PA}(\cdot)$ and $f_{DA}(\cdot)$ yields the total aggregated latency $f_A(\cdot)$.

\textbf{Modeling Dense computation latency.}
%
Dense modules employ stateless computations, meaning their latency depends solely on the number of query tokens, denoted as $n(t)$. For decoding requests, $n(t)$ corresponds to the final token, while for prefill requests, it includes all prompt tokens. However, as demonstrated in Fig.~\ref{fig:modeling}(c), these modules exhibit non-linear latency scaling relative to $n(t)$, accompanied by spike events. This behavior arises because GPUs must allocate new thread blocks when input sizes exceed the hardware’s tile capacity. Threads within these blocks execute in lockstep, even if some redundantly process data (pseudowork). Between spike events, the latency of Dense modules increases linearly with $n(t)$, primarily due to the variable overhead of transferring input tensors from GPU global memory to shared SRAM.


\begin{algorithm}
\caption{The interpolation-based algorithm for the lantecy modeling of Dense modules}
\label{alg:modeling}
\begin{algorithmic}[1]
\small
\Function{modeling}{$min, max, threshold$}

	\State $min\_latency \gets latency(min)$
	\State $max\_latency \gets latency(max)$
	\If{$max\_latency - min\_latency <= threshold$}
		\State $\texttt{Interpolate\ latency\ between\ }min\ \texttt{and}\ max$
	\Else
		\State $mid \gets mean(min + max)$
		\State $mid\_latency \gets latency(mid)$
		\State $min\_part \gets \textsc{MODELING}(min, mid, threshold)$
		\State $max\_part \gets \textsc{MODELING}(mid+1, max, threshold)$
		\State $\texttt{Aggregate modeling result from\ } min\_part\ \texttt{and}\ max\_part$
	\EndIf
\EndFunction

\end{algorithmic}
\end{algorithm}

Motivated by these observations, we propose Alg.~\ref{alg:modeling}, a latency modeling algorithm designed to balance high fidelity and computational efficiency. The algorithm leverages a divide-and-conquer strategy: if a flat latency region is detected (line 4), it interpolates latency values within that interval; otherwise, it recursively identifies spikes in progressively smaller sub-regions (lines 9-10). The threshold for distinguishing flat regions from spikes is dynamically determined as the latency difference between input sizes 1 and 16 for each module. This approach achieves a complexity of $\mathcal{O}(\log n)$, where $n$ represents the maximum supported number of query tokens. Crucially, the algorithm inherently accounts for collective communication overhead, which scales linearly with input size under fixed parallelism conditions.

\textbf{Modeling hybrid parallelism.} Contemporary LLM deployments typically use hybrid parallelism—combining tensor and pipeline parallel techniques—to maximize serving capacity across hierarchical network topologies~\cite{megatron,megatronlm,gpipe}. Our modeling approach integrates seamlessly with the parallelism adopted by users by explicitly accounting for network overhead $\gamma(\cdot)$. In particular, we model the collective communication latency for tensor parallelism as $\gamma_T(\cdot)$ and the peer-to-peer transmission latency for pipeline parallelism as $\gamma_P(\cdot)$, both parameterized by $n(t)$, the number of tokens involved, following the Alpha–Beta communication model~\cite{alphabeta}.



\subsubsection{Scheduling policy}
The Online Scheduler first determines a delicate scheduling order among BE and LS requests performing computations at prefill and decoding phases. 
Ensuring SLO guarantee is of top priority in a cluster, the scheduler prioritizes LS requests over the BE counterparts. Additionally, it incorporates an admission control mechanism to dynamically determine the number of LS prefill requests that can be admitted based on the cluster load, preventing SLO violations.

For all pending and ongoing requests, the Scheduler maintains the following scheduling order across different types: {\color{orange}\ding{182}} LS decoding, {\color{orange}\ding{183}} LS chunk prefill, {\color{orange}\ding{184}} BE chunk prefill, and {\color{orange}\ding{185}} BE decoding. Decoding requests are prioritized over prefill requests for LS services to ensure SLO guarantees on token generation speed. Conversely, BE prefill requests take precedence over BE decoding requests to improve service throughput.
Requests within the same type are executed using a first-come-first-serve policy. During the inference, the Online Scheduler monitors and makes scheduling decisions based on state parameters $c_{PA}(t)$, $c_{DA}(t)$, $g(t)$, and $n(t)$.

\subsubsection{Admission control for LS requests} 
In a resource-constrained cluster, it becomes unfeasible to maintain SLO guarantees for every incoming request during periods of bursty arrivals. In such cases, early rejection is anticipated if a violation is likely, rather than investing time in waiting for execution~\cite{loongserve}. Online Scheduler enforces admission control mechanisms~\cite{kerveros,reservation,borg} for LS requests to address this issue. It aims to reduce unnecessary prefill queuing during peak periods and ensure the fulfillment of prefill SLO requirements. Specifically, with each arrived LS request $k$, Online Scheduler admits it by evaluating whether the total queuing and prefilling time is within the prefill SLO constraint $\mathcal{S}_p$:
\begin{equation*}
 f_{PA}\big( c_{PA}(t) \big) + f_{DA}\big( c_{DA}(t), g(t) \big) + f_D\big( n(t) \big) \leq \mathcal{S}_p \big/ d - {\gamma\big(n(t)\big)} .
\label{eq:evict}
\end{equation*}
Here, $c_{PA}(t)=\sum_{j\in\mathbb{P}(t) \cup k} \sum_{i=l_j + 1}^{p_j} i$, $g(t)=|\mathbb{D}(t)|+|\mathbb{P}(t)|+1$, $c_{DA}(t)=\sum_{j\in \mathbb{D}(t) \cup \mathbb{P}(t) \cup k} l_j+1$, and $n(t)=\sum_{j\in\mathbb{P}(t) \cup k} (l_j - p_j) + |\mathbb{D}(t)|$,  where $\mathbb{P}(t)$ and $\mathbb{D}(t)$ denote the set of incompleted prefill and decoding LS requests respectively, $p_j$ and $l_j$ are the total prompt length and the context length already prefilled for each request $j$, and $d$ is the number of model layers. Notably, this equation offers an explicit  TTFT quantification for each newly arrived request.

\subsubsection{Chunk prefill control} 
\label{ssub:chunk_prefill_control}
After admitting the prefill requests, Online Scheduler determines how many LS prefill loads can be accommodated alongside ongoing LS decoding requests to ensure compliance with the specified SLO. Specifically, for an LS prefill request $j$, it limits the number of tokens $q_j(t)\in\mathbb{Z}^+$ chunk-prefilled at the current iteration, aiming to satisfy the decoding constraint. This can be formulated as an optimization problem:
\begin{subequations}
\begin{align*}
\mbox{max} & \ q_j(t)  \\    
\mbox{s.t.} & \  f_{PA}\big( c_{PA}'(t)\big)  \text{+} f_{DA}\big( c_{DA}(t), g(t) \big)  \text{+} f_D\big( n'(t)\big)\ \text{$\leq$}\ \mathcal{S}_d \big/ d -  {\gamma\big(n(t)\big)} , \\
 & \ q_j(t)  \text{+} l_j \leq p_j.
\end{align*}
\label{eq:prefill_control}
\end{subequations}
Here, $c_{PA}'(t)= c_{PA}(t)+\sum_{i=l_j\text{+}1}^{l_j\text{+}q_j(t)} i$, and $n'(t)=n(t)+q_j(t)$. Additionally, $\mathcal{S}_d$ represents the decoding SLO, while $c_{DA}(t)$, $c_{PA}(t)$ and $n(t)$ are updated to reflect the new computation loads once request $j$ is admitted into the batch.  This optimization problem can be efficiently solved using binary search, due to the monotonically increasing nature of the latency as additional loads are introduced.

When scheduling a BE prefill request, Online Scheduler also determines the chunk-prefill load using the above method with a stricter decoding SLO constraint. Specifically, it replaces $\max\{0, \mathcal{S}_d/d - \omega\}$ by $\mathcal{S}_d/d$ if there are any available Attention results on the CPU Attention output queue. Here, $\omega$ represents the additional piggyback overhead. Meanwhile, BE chunk-prefill load is constrained by the need to satisfy the LS prefill SLO.

\subsubsection{BE decoding control}
After scheduling LS requests and BE prefill loads, the Scheduler seeks to accommodate additional BE decoding requests if there is available capacity on GPUs. Specifically, it determines whether a decoding request $j$ of context length $l_j$ can be executed on the GPU by verifying that the decoding SLO for LS services will not be violated. This check involves simulating the impact of the request on system resources by updating the state parameters:
\vspace{-1em}
\begin{equation*}
 c'_{DA}(t)=c_{DA}(t)+l_j;\ n'(t)=n(t)+1.
\vspace{-0.5em}
\label{eq:gpucheck}
\end{equation*}
The Online Scheduler also reserves room for piggybacking computations by incorporating $\max\{0, \mathcal{S}_d/d - \omega\}$ in the constraint. Additionally, BE decoding requests that cannot be scheduled on GPUs will be offloaded to CPUs for Attention computation.  Conversely, if there are not enough BE decoding requests on the GPUs and the SLO is still maintained, the Scheduler will notify the CPU Attention Manager to swap BE decoding requests back to GPUs, provided there is available GPU memory for additional KV caches.

\subsubsection{Piggybacking control}
\label{sec:piggybacking_budget}
Finally, the Online Scheduler dynamically regulates Attention Piggybacking loads to maintain SLO compliance amidst fluctuating LLM inference workloads. This is achieved through layer-wise batching control, which constrains the number of piggybacked requests ($p_l(t)$) per model layer $l$. To handle dynamically arriving BE piggybacked workloads, the system implements a greedy layer-wise admission strategy. Specifically, the Scheduler incrementally admits BE requests starting from the lowest model layer (ascending order) until SLO thresholds are satisfied. This approach prevents starvation of BE requests in higher layers because: (1) new BE loads are not continuously admitted due to prefill-rate limitations imposed by chunked prefill slots from latency-sensitive (LS) workloads, and (2) requests admitted at lower layers progressively shift upward in subsequent processing cycles.

\begin{figure}
\centering
\includegraphics[width=0.6\linewidth]{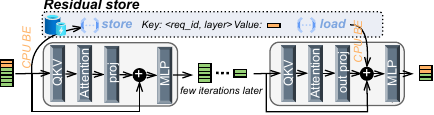}
\vspace{-0.5em}
\caption{The residual store that manages residual tensors for offloaded BE requests for inference correctness.}
\label{fig:residualstore}
\vspace{-1em}
\end{figure}

\section{System Implementation}

\textbf{The implementation of CPU Attention.} Parallelizing the Attention computation among multiple CPU cores and hosts is beneficial to the BE service performance. We adopt the \texttt{OpenMP}~\cite{openmp} library to spawn multiple threads across different CPU cores, where each thread is responsible for performing partial computation loads and aggregating intermediate results. Within each core, we employ Intel® Advanced Vector Extensions (e.g., AVX) instructions to vectorize the Attention computation. When multiple GPUs are attached to a single host, we dedicate an equal share of CPU resources (cores and memory) to each GPU worker. To isolate their interactions with the CPU, each worker is assigned private input and output queues for CPU Attention. The corresponding Attention computation is then performed independently within each worker's allocated resource partition.

\textbf{The implementation of CPU Attention queue.} To mitigate the interference caused by writing and loading overhead on the Attention input and output queues, we implement these two queues in GPU memory as special tensors with additional head and tail pointers to maintain their functionality. This approach prevents blocking in the GPU stream due to costly CPU-GPU transmissions. Only a small amount of additional memory is used as they are solely used to store temporary activations (i.e., $q$, $k$, $v$, and Attention result of one layer). With \texttt{CUDA IPC} and \texttt{MPS} service, CPU process can load and write the queues without blocking the workflow on GPU.

\textbf{Distributed CPU Attention.} Leveraging abundant CPU-only servers in datacenters, we also implement a mechanism called hierarchical CPU Attention to fully utilize the cluster resources. Specifically, the offloaded BE requests are first served on the local host as long as the memory is sufficient. After the local memory is fully occupied, the remaining requests are evenly distributed to remote CPU hosts. KV cache migration and remote Attention functionalities are implemented on top of \texttt{RAY} framework~\cite{ray}.

\textbf{Residual management.} 
Modern LLM models incorporate residual connections~\cite{resnet} in each layer, which span across modules and add complexity to the implementation of the Attention Piggybacking mechanism. In particular, the residual of each request lies on the critical path at every model layer. To preserve computational correctness for BE requests after CPU Attention, we implement a residual store (see Fig.~\ref{fig:residualstore}), which manages the storage and retrieval of residual values throughout the offloaded inference process. For instance, when a BE request has its KV cache offloaded to CPU memory, its residual is saved to the residual store before entering the \texttt{QKV} module, indexed by \texttt{req\_id} and \texttt{layer}. After several iterations, the Attention result is returned to the GPU at the same layer. The previously saved residual tensor is then retrieved from the store using the same identifiers and combined with the output of the \texttt{out\_proj} module to produce the correct result.

\textbf{Profiler and Online Scheduler.} To accurately develop latency models as discussed in~\cref{sec:modeling}, we utilize the linear regression functionality provided by \texttt{sklearn}~\cite{sklearn} to capture the coefficients and intercepts of linear functions. Moreover, we rewrite the scheduler module in vLLM to enable our hybrid load co-serving design.

\section{Experimental Evaluation}
\subsection{Experiment Setup}
\subsubsection{Testbed and workloads} 
We primarily conducted our experiments in a cluster that includes a GPU server equipped with four A100 80GB GPUs, along with four CPU-only servers. The GPUs within the host are interconnected via PCIe channels. Each GPU and CPU server is powered by an Intel® Xeon® Gold 6342 CPU and features 400GB of available RAM. The CPU in the GPU server also participates in Attention computation, alongside the CPU-only servers. All servers are interconnected by default through 100 Gbps RoCE links. Additionally, to evaluate performance across diverse network environments, we report results from specific experiments conducted under a 10 Gbps LAN.

We evaluated OmniServe (OS) on two model architectures: Yi-34B (deployed with 2-way tensor parallelism) and Llama-2-70B (4-way tensor parallelism), demonstrating its general applicability across diverse model scales. We also reported experimental results from two environments: a single GPU server (1G) and a cluster of one GPU server plus four CPU servers (1G4C). Each GPU server contains four GPUs. Unless otherwise specified, all reported results use the latter (1G4C) as the default configuration. All computations used the BF16 precision format on both GPU and CPU instances. The following services were benchmarked:

\textbf{LS service - Online chatbot:} Modeled after real-world conversational workloads, LS requests simulate an interactive chatbot using query-length distributions derived from ShareGPT~\cite{sharegpt}. Requests were continuously submitted to the system at a fixed rate with Poisson arrival patterns by default. We also evaluated the SLO attainment of the LS service under highly dynamic submission rates, discussed in~\cref{sec:slo_attainment_result}.

\textbf{BE service:} BE workloads were generated using two benchmarks: 1) LongBench-v2~\cite{longbench}: Average input/output lengths are 8,952/136 tokens, with a maximum length of 12K tokens. 2) DailyMails~\cite{cnndailymail}: Average input/output lengths is 1,964/397 tokens. The cluster was configured to simulate a BE request load by replaying a submission pattern from the Azure Public Datasets~\cite{dynamollm}, where 182.6 requests are submitted per minute on average.

\subsubsection{Performance metrics} 
We used SLO attainment and token generation throughput as the primary evaluation metrics for LS and BE services, respectively. For LS services, the attainment rate is assessed based on both TTFT and TPOT. We defined fixed TPOT and TTFT constraints similar to previous works~\cite{distserve}. For BE services, we focused on token generation rates during both the prefill and decode phases. Without specific mention, for 34B and 70B models, the TTFT SLOs for LS service is set to 2s and 3s, while the TPOT SLOs are set to 0.2s and 0.25s, respectively. All experiments were conducted over a period of 30 minutes.

\subsubsection{Baselines} We adopted the following baselines:

\textbf{Baseline A: Llumnix~\cite{llumnix} on GPU + vLLM~\cite{pagedattention} on CPU.}
We adopt the Llumnix~\cite{llumnix} system to host the hybrid inference loads on GPUs, which utilizes a memory-centric control policy to achieve performance isolation. Moreover, to fully utilize the CPU resources in clusters for a fair comparison with OmniServe, BE requests that could not be accommodated on GPU instances were offloaded to the CPU-hosted vLLM~\cite{pagedattention} instances and their inference process would be computed on CPUs.

\textbf{Baseline B: NEO~\cite{neo}.}
We adapt NEO that leverages both CPU and GPU for latency-oriented LLM inference. NEO identifies the decoding phase's Attention computation as being relatively lightweight and offloads it entirely to the CPU. To harness both devices, it employs a pipeline pattern: while the CPU processes the Attention for one micro-batch of requests, the GPU concurrently executes the non-Attention modules for another. However, NEO's performance is ultimately constrained by its reliance on the CPU for all Attention computation, making it vulnerable to bottlenecks in CPU processing power and PCIe bandwidth, especially under strict SLOs. For a fair comparison, we enhanced NEO with a latency control mechanism similar to OmniServe's to prevent SLO violations.

\textbf{Baseline C: Sarathi-Serve \protect\cite{taming}.} We also adapt Sarathi-Serve as our SLO-optimal baseline. This system runs computations solely on the GPU, eliminating potential SLO degradation caused by CPU offloading. It consistently prioritizes LS requests, with overflowed BE requests queued for available GPU slots. Experimental results from this baseline establish the upper bound for SLO achievement without CPU assistance across diverse LS submission patterns.

\begin{figure}
\centering
\includegraphics[width=0.56\linewidth]{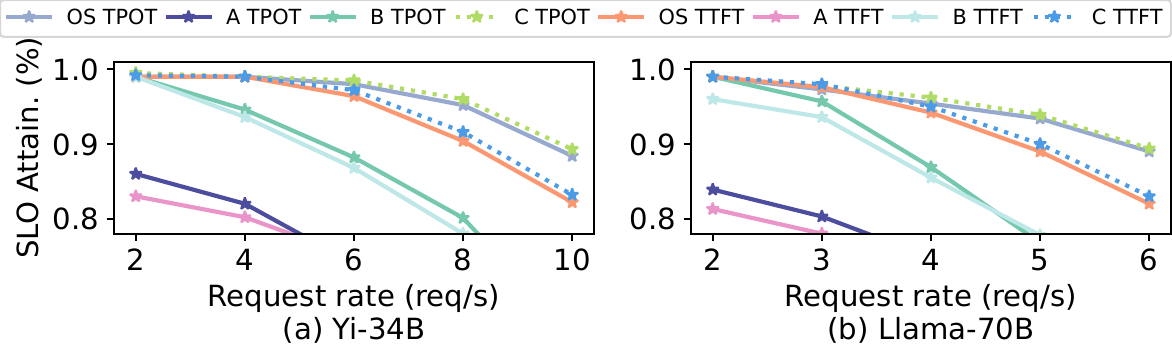}
\vspace{-0.5em}
\caption{SLO attainment across LS arrival rates with LongBench-v2 dataset. 1 GPU  and 4 CPU hosts are used.}
\label{fig:req_rate_longbench_1G4C}
\vspace{-1em}
\end{figure}

\begin{figure}
\centering
\includegraphics[width=0.56\linewidth]{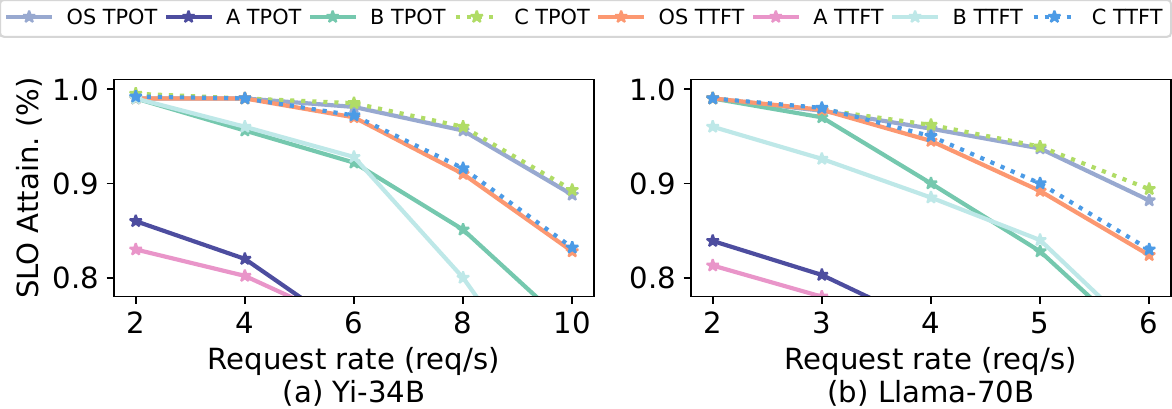}
\vspace{-0.5em}
\caption{SLO attainment across LS arrival rates with LongBench-v2 dataset. Only 1 GPU server is used.}
\label{fig:req_rate_longbench_1G}
\vspace{-1em}
\end{figure}

\begin{figure}
\centering
\includegraphics[width=0.56\linewidth]{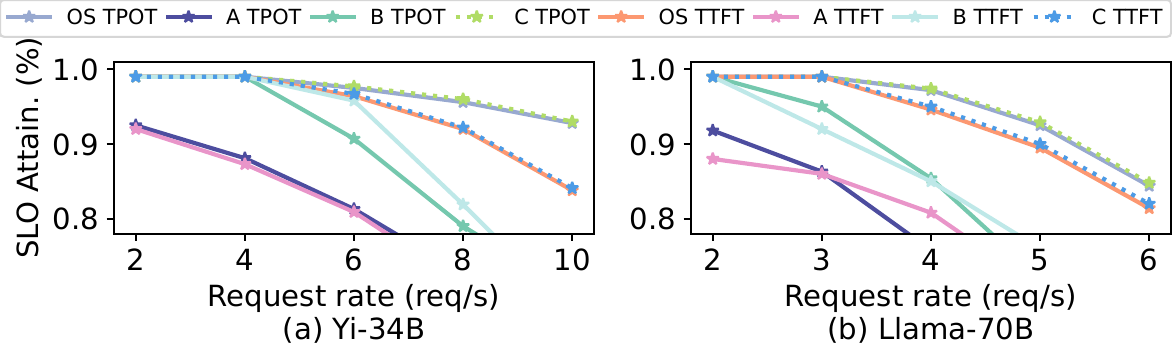}
\vspace{-0.5em}
\caption{SLO attainment across LS arrival rates with DailyMails dataset. Only 1 GPU server is used.}
\label{fig:req_rate_dailymails}
\vspace{-1em}
\end{figure}

\subsection{End-to-end Performance}
We evaluated the end-to-end performance of OmniServe against baselines under various conditions. We selected the request arrival rate and SLO constraints to ensure that all LS requests are admitted to the system without any rejections.

\subsubsection{SLO attainment of LS services} 
\label{sec:slo_attainment_result}

In Fig.~\ref{fig:req_rate_longbench_1G4C} and Fig.~\ref{fig:req_rate_longbench_1G}, we examined the SLO achievement of the systems across varying arrival rates under two hardware configurations, i.e., 1G4C and 1G. The BE requests are generated from the LongBench-v2 dataset. It is evident that the SLO achievement rates for all baselines decrease as the LS request arrival rate increases. This decline occurs because the upper limit of serving capacity is constrained by the allocated GPU resources. As the number of LS requests increases, competing for resources with BE loads, Llumnix experiences significant performance degradation. This is primarily because its memory-based allocation fails to maintain the latency objectives, as factors such as request length and inference phase also play critical roles. The NEO baselines also struggle to achieve high SLO rates in this scenario. Although NEO guarantees that latency remains below specified thresholds, its design for CPU-GPU interaction limits the capability to handle a larger number of requests. Notably, NEO does not differentiate between LS and BE workflows, resulting in all LS requests' Attention being processed on the CPU. Consequently, both PCIe bandwidth and limited CPU computational power can become bottlenecks for LS services. For this reason, its SLO attainment is even worse given more CPUs due to the increased communication overhead. In contrast, OmniServe consistently sustains as high SLO attainment rates as Sarathi-Serve in both environments, where up to only 0.6\% SLO attainment degradation is introduced. This is attributed to OmniServe's nuanced latency control strategy, which considers multiple factors, including token count and context length. Additionally, it decouples the servicing of LS and BE requests, preventing LS services from being bottlenecked by CPU resource limitations. Moreover, in Fig.~\ref{fig:req_rate_dailymails}, OmniServe achieves up to $1.42\times$ higher SLO attainment with BE requests from the DailyMails dataset, proving its adaptability to a diverse computational intensity imposed by BE services.

\begin{figure}[t]
\centering
\includegraphics[width=0.56\linewidth]{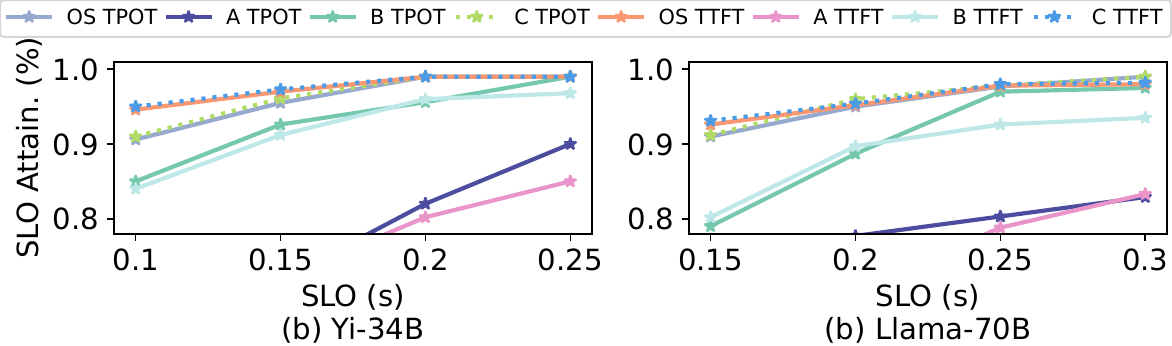}
\vspace{-.5em}
\caption{SLO attainment across SLO constraints with LongBench-v2 dataset. Only 1 GPU server is used.}
\label{fig:slo_rate}
\vspace{-1em}
\end{figure}

Moreover, we evaluated SLO performance under varying latency requirements, with arrival rates set to 4 req/s and 3 req/s for the 34B and 70B models, respectively. As depicted in Fig.~\ref{fig:slo_rate}, the SLO attainment rate across all baselines improves as the latency constraint is relaxed, given a fixed allocation of GPU resources. However, Llumnix struggles to maintain high SLO attainment under stringent TPOT requirements, particularly when interfered with by BE requests. For instance, when the SLO is set to 0.15s, its TPOT attainment rate for the Llama-70B model drops to $62\%$, whereas OmniServe maintains a $91.6\%$ SLO attainment rate and presents 1.48$\times$ improvement. This performance gap occurs because stringent SLOs are more vulnerable to latency exacerbation caused by admitting BE serving loads. A similar trend is observed with the DailyMail dataset.

\begin{figure}[t]
\centering
\includegraphics[width=0.56\linewidth]{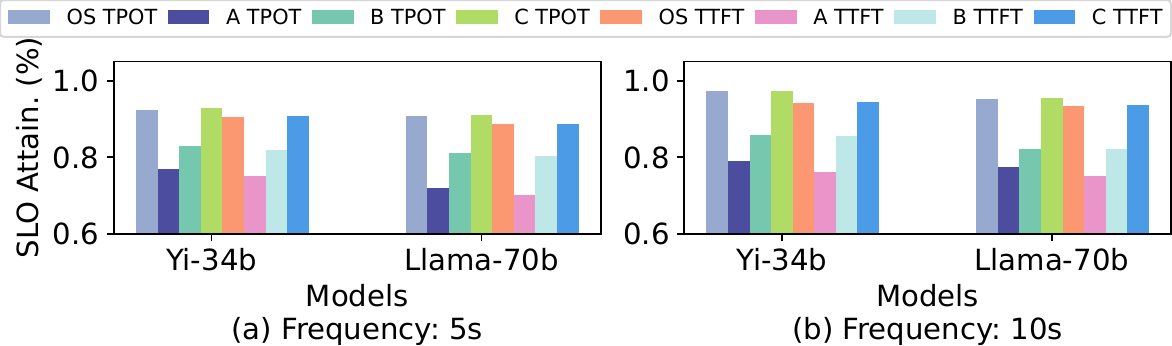}
 \vspace{-1em}
\caption{SLO attainment under the bursty request arrival test. The submission rate randomly varies from 1 to 8 for Yi-34B and from 1 to 5 for Llama-70B, respectively. The interval of changing the submission rates for two models are 5s and 10s. Evaluations are conducted on the 1G hardware setting.}
\label{fig:ls_slo_dynamic}
\vspace{-1em}
\end{figure}

Furthermore, we evaluated SLO attainment under dynamic workload arrival patterns based on a real-world LS submission trace with varying intensity over time. The request submission rate was modified at two different frequencies (5s and 10s), with each modification event assigning a random rate between 1–8 req/s for the Yi-34B model and 1–5 req/s for the Llama-70B model. As shown in Fig.~\ref{fig:ls_slo_dynamic}, OmniServe consistently outperforms Llumnix and NEO, achieving up to $1.23\times$ and $1.13\times$ higher SLO attainment, respectively. Moreover, OmniServe always maintains nearly identical SLO attainment with Sarathi-Serve, proving that there is no performance sacrifice under bursty loads. It stems not only from asynchronous CPU-GPU coordination but also from the cache management mechanism in~\cref{sec:cache_management}. By minimizing expensive KV cache migration under fluctuating loads, it helps maintain high inference efficiency.

\begin{figure}
\centering
\includegraphics[width=0.56\linewidth]{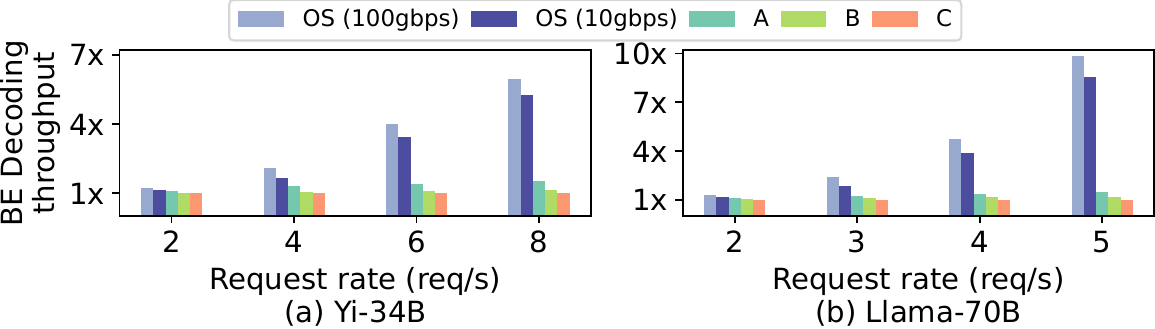}
\vspace{-1em}
\caption{Decoding throughput of BE service using LongBench-v2 dataset. 1 GPU and 4 CPU hosts are used.}
\label{fig:be_throughput_longbench_1G4C}
\vspace{-1.2em}
\end{figure}

\begin{figure}
\centering
\includegraphics[width=0.56\linewidth]{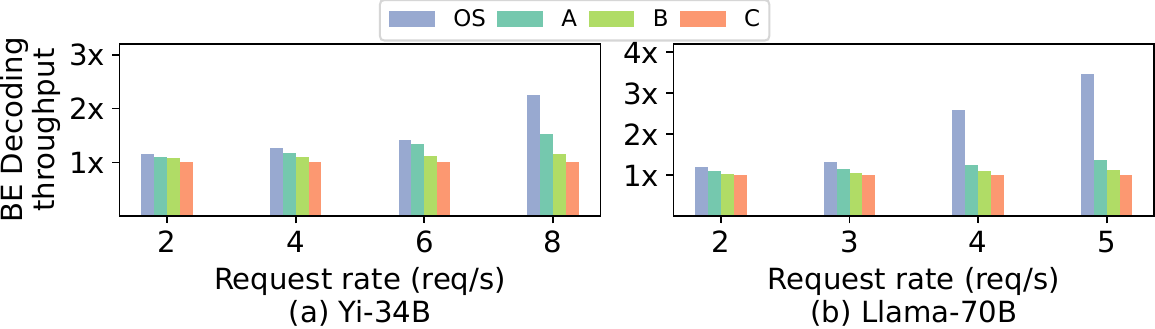}
\vspace{-1em}
\caption{Decoding throughput of BE service using LongBench-v2 dataset. Only 1 GPU server is used.}
\label{fig:be_throughput_longbench_1G}
\vspace{-1.2em}
\end{figure}

\subsubsection{BE serving throughput} 
\begin{figure}
\centering
\includegraphics[width=0.56\linewidth]{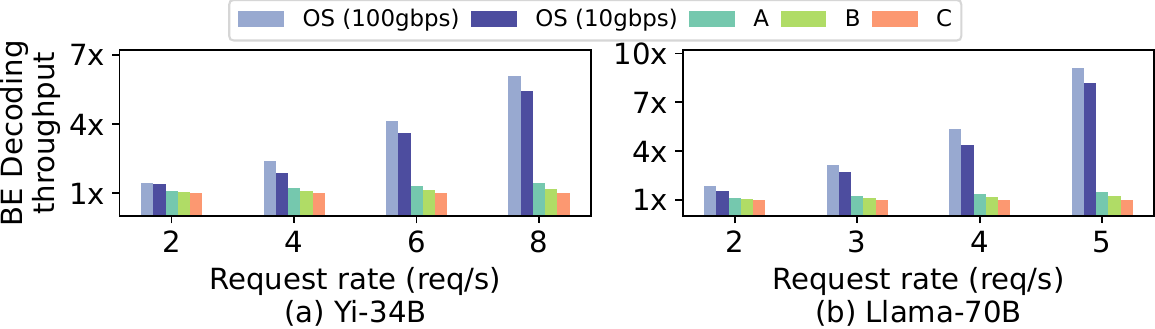}
\vspace{-0.5em}
\caption{Decoding throughput for BE requests under DailyMails dataset. 1 GPU and 4 CPU hosts are used.}
\label{fig:be_throughput_dailymails}
\vspace{-1em}
\end{figure}

This section evaluates BE throughput across baseline systems. To comprehensively evaluate OmniServe's efficiency, we conducted benchmark tests under the following hardware configurations: (1) a single GPU server (1G), and (2) one GPU server plus four CPU servers (1G4C). In the latter configuration, we examined two network setups, i.e., 100 Gbps and 10 Gbps. Across these environments, the baseline Sarathi-Serve always yields the lowest BE throughput, as it cannot leverage the CPU power. The BE performance under NEO is bottlenecked by strict SLO requirements, which restrict the BE computation on the CPU. In Fig.~\ref{fig:be_throughput_longbench_1G4C} and Fig.~\ref{fig:be_throughput_longbench_1G}, OmniServe consistently outperforms baselines on the LongBench-v2 benchmark. Under light LS workloads with ample GPU resources that can be leveraged by the BE service, it achieves a modest $1.2\times$ throughput improvement, as the GPU's computational capacity reduces dependence on CPU-assisted processing. However, as LS workloads intensify and GPU resources for BE requests diminish, efficient CPU utilization becomes critical. While baseline systems struggle with inefficient CPU-based computation—creating significant bottlenecks—OmniServe's coordinated GPU-CPU orchestration delivers a $9.85\times$ throughput advantage under heavy load. This performance advantage persists in low-bandwidth environments: by transmitting only intermediate tensors, our lightweight Attention Piggybacking design minimizes communication overhead. Moreover, even under the configuration where only one CPU in the GPU server is involved, OmniServe can still achieve $3.47\times$ improvement, proving its pervasive efficiency in CPU-limited conditions. The benefits are also pronounced for BE requests under the DailyMail benchmark in Fig.~\ref{fig:be_throughput_dailymails}, where BE requests have relatively shorter context lengths. Although baselines can serve more BE requests on the GPU in this case, OmniServe can still achieve up to $9.1\times$ improvement, proving its applicability in practical scenarios with diverse characteristics.

\subsection{Effectiveness of Attention Piggybacking}
\label{piggyback-attention-experiment}
In this section, we explore the benefits and impacts of the Attention Piggybacking mechanism. For BE services, experiments were conducted using the LongBench-v2 dataset, while LS workloads were tested with request rates of 4 reqs/s and 3 reqs/s for the 34B and 70B parameter models.

\begin{figure}
\begin{minipage}{0.22\linewidth}
\includegraphics[width=0.99\linewidth]{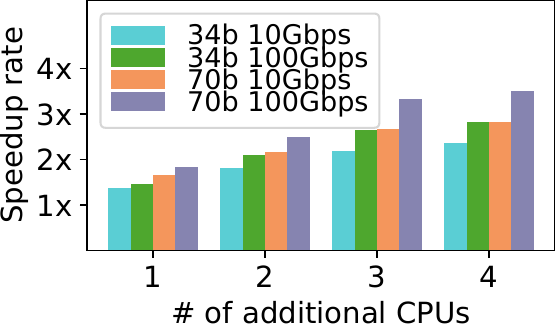}
\subcaption{BE throughput}
\end{minipage}
\begin{minipage}{0.22\linewidth}
\includegraphics[width=0.99\linewidth]{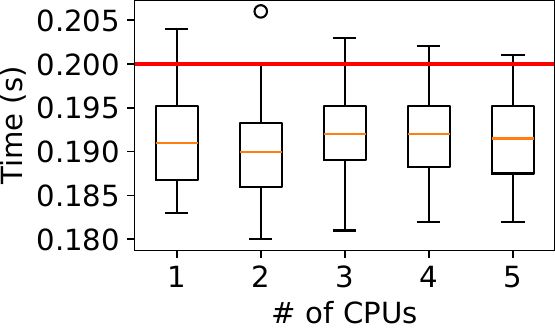}
\subcaption{LS TPOT}
\end{minipage}
\vspace{-0.5em}
\caption{The benefit and impact of admitting more CPUs in Attention Piggybacking on BE and LS services.}
\vspace{-1em}
\label{fig:piggybacking_impact}
\end{figure}

First, we conducted experiments to examine the benefit of leveraging a varying number of CPUs in clusters. In Fig.~\ref{fig:piggybacking_impact}(a), we observe a significant and consistent improvement in BE throughput under OmniServe when utilizing more CPUs, compared to the throughput with an attached CPU in the GPU host only. Thanks to the distributed Attention mechanism, we can fully utilize the CPU computational power in cluster. Since the communication overhead of transferring intermediate results for BE requests is significantly less than the Attention computation time, the communication overhead can be effectively hidden by the computation, yielding near-linear throughput improvement. To this end, up to $3.43\times$ speedup is witnessed given four additional CPUs.

Second, we investigated the impact of involving more CPUs to SLO maintenance. As illustrated in Fig.~\ref{fig:piggybacking_impact}(b), the median token generation latency remains nearly constant as the number of CPUs increases. Furthermore, the maximum latency consistently aligns with the decoding TPOT SLO. This demonstrates that the piggybacking control does not adversely affect LS services, thanks to the asynchronous communication and computation design, as well as the precise control over the number of BE requests piggybacked for the computation of Dense modules at each layer. 

\begin{figure}
\begin{minipage}{0.2\linewidth}
\includegraphics[width=0.99\linewidth]{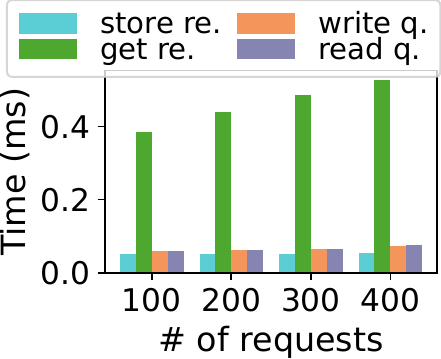}
\subcaption{Piggybacking operation overhead}
\end{minipage}
\begin{minipage}{0.18\linewidth}
\includegraphics[width=0.99\linewidth]{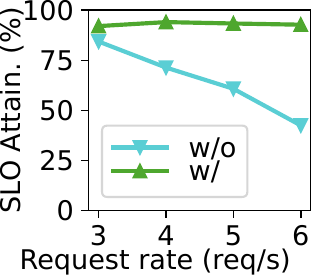}
\subcaption{LS TTFT SLO attainment rate}
\end{minipage}
\begin{minipage}{0.18\linewidth}
\includegraphics[width=0.99\linewidth]{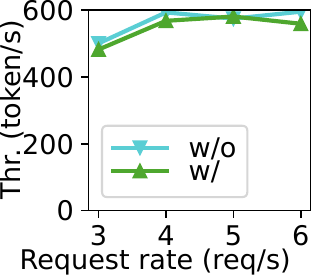}
\subcaption{LS decode under admission control}
\end{minipage}
\vspace{-0.5em}
\caption{(a) Illustration of OmniServe's Attention piggybacking overhead, including residual and queue operations; (b) and (c) jointly reveals the admission control design maintains prefill SLO without sacrificing decoding throughput. 
}
\vspace{-1em}
\label{fig:micro_eval}
\end{figure}

Third, we evaluated the implementation overhead of the Attention Piggybacking mechanism. Specifically, we quantified the latency introduced by auxiliary operations in the inference pipeline across layers, varying the number of piggybacked BE requests. In Fig.~\ref{fig:micro_eval}(a), these operations incurred negligible overhead: queue read/write operations and residual storage required $\leq$75 µs even for 400 concurrent requests, owing to efficient contiguous memory access. However, residual loading from the residual store tensor—triggered when CPU-generated Attention results return out of sequence—introduces non-contiguous memory access, resulting in higher latency (approximately 0.5 ms for 400 requests). Critically, this cost remains marginal compared to the inference latency, as the piggybacking design inherently limits its frequency: Attention results are deferred by at least one iteration before residual loading occurs, ensuring such operations remain infrequent.

Finally, we conducted experiments to study the piggybacking overhead under varying numbers of CPUs. Specifically, we deployed Yi-34B model in a GPU host, and allocated 1, 2, 3, and 4 additional CPUs, respectively. For each configuration, we compared the SLO attainment rates with and without the BE service using the LongBench-v2 dataset. By this comparison, we can investigate the performance degradation given the piggybacking mechanism for BE service. Results show that the SLO attainment is only downgraded by $0.36\%$, $0.34\%$, $0.28\%$, and $0.31\%$ under these settings, proving consistently low overhead regardless of the number of CPUs allocated.

\subsection{Ablation Study}

\subsubsection{Modeling accuracy} 
We investigated the modeling accuracy of token generation latency. The Attention latency model is built on profiling 100 data samples. We evaluated the modeling accuracy of \textsc{OmniServe} across multiple parallel configurations using eight GPUs, collecting 1,000 samples for each test. As detailed in Table.~\ref{table:latency_accuracy}, the average accuracy for the Yi-34B model reaches 95.7\%. For the larger Llama-2-70B model, which is more susceptible to inter-GPU communication overhead, our approach-which inherently profiles this overhead-maintains a high average accuracy of 94.5\%. Moreover, the P90 accuracy—defined as the 90th percentile value in descending order for each configuration, measuring performance consistency—remains high across all tested setups, with values no lower than 93.6\% and 92.7\% for Yi-34B and Llama-2-70B, respectively, in any parallel configuration.

\begin{table}[]
\caption{The accuracy of OmniServe's latency model under different parallel configurations. *Average and 90th percentile values across settings (example: PP2\&TP4 = 2 pipeline stages, each with TP degree of 4)*}
\vspace{-.8em}
\small
\begin{tabular}{ccccc}
\hline
\multicolumn{1}{|c|}{}              & \multicolumn{1}{c|}{PP8\&TP1} & \multicolumn{1}{c|}{PP4\&TP2}  & \multicolumn{1}{c|}{PP2\&TP4}  & \multicolumn{1}{c|}{PP1\&TP8} \\ \hline
\multicolumn{1}{|c|}{Yi-34B} & \multicolumn{1}{c|}{94\%, 92.2\%}  & \multicolumn{1}{c|}{95.2\%, 93.6\%} & \multicolumn{1}{c|}{94.9\%, 93\%} & \multicolumn{1}{c|}{95.7\%, 93.3\%}  \\ \hline          
\multicolumn{1}{|c|}{Llama-70B} & \multicolumn{1}{c|}{94.3\%, 92.7\%}  & \multicolumn{1}{c|}{93.8\%, 91.5\%} & \multicolumn{1}{c|}{93.2\%, 91\%} & \multicolumn{1}{c|}{94.5\%, 92.1\%}  \\ \hline            
\end{tabular}
\label{table:latency_accuracy}
\vspace{-1em}
\end{table}

\subsubsection{Admission control} 
We evaluated the effectiveness of OmniServe’s admission control in optimizing performance for LS services. As illustrated in Fig.~\ref{fig:micro_eval}(b), without admission control, SLO attainment rates drop sharply as request rates increase, falling to critical levels under high load. In contrast, with admission control enabled, OmniServe sustains a 94.1\% TTFT SLO attainment rate, even under aggressive workloads. This improvement stems from mitigating the adverse effects of indiscriminately admitting requests beyond system capacity—specifically, excessive prefill queuing delays and unmanageable KV cache contention. By judiciously regulating admissions, OmniServe achieves up to 43.3\% higher prefill SLO compliance compared to baseline approaches.

Notably, this performance gain does not sacrifice LS serving throughput. As shown in Fig.~\ref{fig:micro_eval}(c), LS decoding throughput with admission control remains nearly identical to unregulated scenarios, peaking at a marginal 6\% difference. This indicates that OmniServe efficiently prioritizes high-priority requests without under-utilizing GPU resources. Furthermore, the scheduling overhead introduced by admission control is minimal, remaining below 1ms even at peak loads (e.g., 8 requests per second). This overhead is negligible relative to the time required to generate a single token, ensuring the design’s practicality for real-time inference.

\subsubsection{Operational overhead}

Since we leverage RAY to implement the cluster-level control plane, we also evaluated the operational overhead caused by this framework. Specifically, we measured the inference latency with and without the RAY framework to isolate its overhead. In the non-RAY baseline, all cross-host control logic (e.g., hardware discovery) was disabled, while the input data and model remained identical for both cases. Experimental results demonstrate that RAY only induces up to $0.98\%$ and $0.9\%$ latency increase for a Yi-34B and Llama-70B model, respectively.

\subsubsection{Performance in PD-disaggregation setting}

We also evaluated OmniServe's performance under a Prefill-Decode disaggregation setup, demonstrating its strong compatibility. In this configuration, we deployed the LS  service across separate prefill and decode instances, each using a tensor-parallel degree of 2. Hybrid serving on both instance types was facilitated by a customized SLO‑aware scheduler, operating in a manner analogous to~\cref{sec:scheduling}: it prioritizes LS requests and then utilizes any remaining computational resources for BE tasks. Additionally, the decode instance employs a piggybacking mechanism to offload part of the BE decoding Attention computation to the CPU.

Using a Llama‑2‑70B model with an LS request rate of 4 req/s and BE workloads from LongBench‑v2, OmniServe achieves up to 1.48$\times$ higher SLO attainment rate and 6.94$\times$ greater BE decoding throughput compared to Llumnix.

\section{Discussion}

In this section, we present potential extensions and optimizations for leveraging OmniServe in LLM hybrid serving scenarios.

\textbf{Supporting inference with multiple priority levels.} This setting has gained significant attention recently in the field of efficient LLM serving~\cite{adaserve, slosserve, hyperflexis, polyserve}. Typically, high-priority requests come with tight SLOs, whereas lower-priority requests are subject to more lenient ones. The core idea behind OmniServe—asynchronous inter-request batching—can be adapted to this setting. Specifically, the computation for low-priority requests can be partially deferred at a given layer and then piggybacked onto the execution of the same layer several iterations later, with the deferral interval determined by the corresponding SLO requirement. This approach reduces token-generation latency for high-priority requests compared to continuous batching, as it alleviates contention for computational resources. Building on this, the scheduling policy introduced in~\cref{sec:scheduling} can be naturally extended to formulate similar optimization problems that determine how many requests of each priority type can be scheduled.

\textbf{Supporting more flexible CPU offloading designs.} 
Recent research has proposed offloading non-Attention modules to the CPU as one such strategy~\cite{lia, KTransformers, heterogen}. Although these offloading strategies were not designed for hybrid LLM workloads with distinct priorities, they can be effectively integrated into OmniServe to improve the performance of BE services in various scenarios. For instance, in KTransformers~\cite{KTransformers}, hybrid serving can be supported by partitioning CPU cores to separately handle MoE and Attention computations. Similarly, when BE workloads involve intensive prefill operations but few offloaded KV caches, OmniServe can leverage partial CPU cores equipped with modern instruction sets—such as Intel’s Advanced Matrix Extensions (AMX)—to accelerate Dense computation within the BE service.

\section{Related Work}
\label{sec:cpu_serving}

\textbf{LLM serving with CPUs.} 
A growing body of work focuses on democratizing LLM serving on CPUs. Llama.cpp~\cite{llamacpp} and vLLM~\cite{pagedattention} support offloading partial layers to the CPU. PowerInfer~\cite{powerinfer} treats CPU memory as external storage, where model parameters and input tensors are transferred between CPU and GPU during the inference process. NEO and FastDecode~\cite{neo, fastdecode} identify the light computation of decoding Attention and leverages CPU for decoding Attention to enhance the LS service. They split the inference batch into multiple sub-batches, enabling simultaneous execution on CPU and GPU while reducing resource idle time. HeteGen~\cite{hetegen} enables efficient small-batch LLM serving in resource-constrained environments where GPUs lack the memory to host complete model parameters. To overcome this limitation, it offloads parameters to CPU memory and dynamically orchestrates computation between the CPU and GPU to minimize inference latency. KTransformer and LIA address the same problem as HeteGen but pursue a different optimization~\cite{lia, KTransformers}. They improve inference efficiency by leveraging the Advanced Matrix Extension (AMX) technique, which provides matrix computation performance comparable to that of a GPU. However, since this specialized hardware is not universally available, the significant performance gap indicated in Table~{\ref{tab:gap}} limits the applicability of their strategies in environments without such support. Conversely, OmniServe focuses on the inference scenario where GPUs can host all model parameters. Moreover, these systems are primarily optimized for low latency and, as a result, largely overlook improvements in resource utilization for BE services.

\textbf{SLO-oriented inference.} Nowadays, a body of SLO-oriented works has been proposed. \cite{taming, distserve, splitwise, exegpt} optimize SLO attainment for LS service on GPU-only platform. \cite{hetegen, neo, KTransformers, lia} help improve SLO achievement for LS service only on top of CPU-GPU hybrid platform. Llumnix~\cite{llumnix} is the first work that colocates LS and BE services, which uses GPU resource only. In contrast, OmniServe stands out to be the first work that sustains SLO achievements and improves BE throughput using CPU and GPU resources.

\textbf{LLM serving with heterogeneous GPU devices.} Recently, resource heterogeneity has become an important issue in datacenter~\cite{heet, oef, fft}, and several systems have been developed for efficient serving in heterogeneous GPU clusters~\cite{helix, hexgen, hetis}, with their delicate technique to balance computational and memory resource usage across GPU devices. However, they focus on parallelizing computation and cache storage loads uniformly (e.g., sharding on tensor- or pipeline-parallelism) and applying synchronous data aggregation. As a result, they are also not well-suited for being adapted to the hybrid serving cases with CPUs and GPUs, considering the significant computational power gap in Dense modules.

\textbf{Generalizing to advanced model architecture.}
LoRA (i.e., Low-Rank Adaptation) architecture enables efficient adaptation of a shared foundational model to diverse service-specific generation requirements~\cite{slora, dlora, punica}. During inference, requests from distinct services are dynamically routed to specialized LoRA weights linked to the model’s QKV modules. By retaining QKV computations on GPUs for BE requests—a resource-efficient design choice—OmniServe achieves seamless compatibility with LoRA-driven optimization strategies. Moreover,
the OmniServe architecture also provides native support for MoE models. Because its design confines all dense module computation to the GPU, MoE optimization strategies like Expert Parallelism are inherently compatible without modification.

\section{Conclusion}
This paper presents OmniServe, a new LLM inference system designed for serving hybrid loads with SLO awareness on CPU-GPU platforms. By adopting an innovative Attention Piggybacking inference mechanism and a delicate batching control policy, OmniServe effectively leverages both CPU and GPU resources to optimize BE services while ensuring SLO guarantees for LS services within data centers. In its inference flow design, both CPUs and GPUs execute BE services through asynchronous communication, significantly reducing the interference introduced into the inference process of LS services. This work also creates an opportunity to harness CPU resources to enhance the serving capacity for a single type of service with extremely high request arrivals.

\begin{acks}
This work was supported in part by the Science and Technology Development Fund of Macau (0041/2025/RIA1, 0074/2025/AMJ), the Multi-Year Research Grant of University of Macau MYRG-GRG2024-00255-FST-UMDF, MYRG-GRG2025-00119-FST), and Alibaba Group through Alibaba Innovative Research Program.
\end{acks}

\bibliography{paper}

\end{document}